\documentstyle[12pt]{article}

\begin{document}

\title{Ruling out a higher spin field solution to the cosmological
constant problem}

\author{
  V.A.~Rubakov and  P.G.~Tinyakov \\
  {\small {\em Institute for Nuclear Research of the Russian 
   Academy of Sciences,}}\\
  {\small {\em 60th October Anniversary prospect 
   7a, Moscow 117312, Russia.}}
  }
\date{}
\maketitle

\begin{abstract}
We consider the modification of Newton's gravity law in Dolgov's
higher spin models \cite{Dolgov:1991zc,Dolgov:1997zg} designed to
compensate the cosmological constant. We find that the effective
Planck mass is unacceptably large in these models. We also point out
that the properties of gravitational waves are entirely different in
these models as compared to general relativity.\vspace{0.5cm}

\noindent
Pacs: 04.20.Cv, 98.80.Cq.
\end{abstract}

\section{Introduction} 

One of the most challenging problems in contemporary physics is the
cosmological constant problem: observationally allowed values of the
vacuum energy density are many orders of magnitude larger than those
natural from particle physics point of view (for a review see, e.g.,
ref.\cite{Weinberg:1989cp}). An attractive approach towards the
solution of this problem would be to introduce a compensating field
which interacts with gravity in such a way as to relax dynamically to
a state with zero or very small net energy density. Attempts to make
use of a scalar field for this purpose have failed so far
\cite{scalar1,scalar2}, so it was suggested by Dolgov
\cite{Dolgov:1991zc,Dolgov:1997zg} that vector or tensor fields may do
the job. The latter mechanism is based on the observation that some
components of higher spin fields may develop an instability if the
cosmological constant is large and the Universe is rapidly
expanding. As a result of the instability, these components grow, and
their contribution to the energy-momentum tensor slows down the
expansion of the Universe thus almost completely compensating the
effect of the cosmological constant. A peculiarity of this mechanism is
that the Lorentz invariance is strongly broken in the gravitational
sector.

In this note we point out that the large fields and strong breaking of
Lorentz invariance, inherent in Dolgov's models, have drastic effects
on gravitational interactions. In particular, Newton's gravity gets
modified in an unacceptable manner. Also, the properties of
gravitational waves become entirely different from the conventional
ones. Hence, the mechanism proposed in
refs.\cite{Dolgov:1991zc,Dolgov:1997zg} is not acceptable
phenomenologically.

\section{Compensation mechanism}

In this paper we will consider in some detail a variant of the
compensation mechanism of ref.\cite{Dolgov:1991zc,Dolgov:1997zg} which
makes use of vector field. The analysis of the models with tensor
fields is very similar, and leads to the same conclusions. The action
proposed in ref.\cite{Dolgov:1997zg} is
\[
S = S_G + S_{vac} + S_A
\] 
where 
\[
S_G = - {M^2\over 16\pi} \int d^4x \sqrt{-g} R,
\]
\[
S_{vac} = \int d^4x \sqrt{-g} \epsilon_{vac},
\]
\[
S_A = {\eta\over 2} \int d^4x \sqrt{-g} D_{\mu}A_{\nu} 
D^{\mu}A^{\nu}.
\]
Here we keep $M$ as a free parameter not necessarily equal to the
Planck mass; $\epsilon_{vac}$ is the vacuum energy density of ordinary
matter (whose effect on the expansion of the Universe is aimed to be
compensated), $A_{\mu}$ is the compensating vector field, and $\eta$
is either $+1$ or $-1$, depending on the sign of $\epsilon_{vac}$. The
vacuum energy density $\epsilon_{vac}$ is expected to be determined by
some particle physics scale; say, the contribution of electroweak
physics is naturally of order 
\[
\epsilon_{vac} \sim (100\;\mbox{GeV})^4. 
\]
If not for the field $A_{\mu}$, the Universe would expand way too
fast. 

A homogeneous spatially flat solution to the field equations with
vanishing spatial components $A_i$ has the following large time
asymptotics,
\begin{equation}
A_0 = t \sqrt{\epsilon_{vac}/2},
\label{A0}
\end{equation}
\begin{equation}
H = {1\over t} ,
\label{H}
\end{equation}
where $H$ is the Hubble parameter. The large value of $\dot{A}_0$
compensates the effect of the vacuum energy density, $\epsilon_{vac}$, so
the Universe evolves slowly at large $t$. This happens because 
\begin{equation}
\dot{A_0}^2 \sim A_0^2 H^2 \sim \epsilon_{vac},
\label{main-eq}
\end{equation}
which is the major ingredient of the compensation mechanism. 

It is clear from eq.(\ref{main-eq}) that the present value of $A_0$ is
very large, 
\[
A_0 \sim {\sqrt{\epsilon_{vac}}\over H}. 
\]
Unless $\epsilon_{vac}$ is less than $(10^{-3}\;\mbox{eV})^4$ (i. e.,
unless no compensation of the cosmological constant is needed), this
value at the present epoch is much greater than the Planck mass, 
\begin{equation}
A_0 \gg M_{Pl}. 
\label{AggM}
\end{equation}
As an example, at $\epsilon_{vac}\sim (100\;\mbox{GeV})^4$ one has
$A_0\sim 10^{46}\;$GeV.

\section{Modified gravity law} 

Let us study perturbations about the background determined by
eqs.(\ref{A0}) and (\ref{H}). We consider time scales and wavelengths
of these perturbations much smaller than the expansion time and Hubble
radius of the Universe, respectively. It is these perturbations that
are relevant to Newton's gravity law and emission of gravitational
waves. Under these conditions the largest parameter entering the
quadratic part of the perturbed action $S_A$ is the present value of
$A_0$.  The time derivatives of $A_0$ and background metrics may be
safely ignored; in particular, background metrics is effectively
Minkowskian. The quadratic part of $S_A$ then takes the following
form,
\[
S^{(2)}_A =     
{\eta\over 2} \int \Bigl[
\partial_{\mu} c_{\nu}\partial^{\mu} c^{\nu} 
- A_0 \partial^{\mu} c^{\nu} (\partial_{\mu} h_{\nu 0} 
+ \partial_{\nu} h_{\mu 0} - \partial_0 h_{\mu\nu} )
\]
\begin{equation}
+ {A_0^2 \over 4} (\partial_{\mu} h_{\nu 0} 
+ \partial_{\nu} h_{\mu 0} - \partial_0 h_{\mu\nu} )
(\partial^{\mu} h^{\nu 0} 
+ \partial^{\nu} h^{\mu 0} - \partial^0 h^{\mu\nu} ) \Bigr]
\label{S2}
\end{equation}
where $c_{\mu}$ and $h_{\mu\nu}$ are perturbations of the vector field
and metrics, respectively. The quadratic parts of $S_G$ and $S_{vac}$
are conventional and we do not write them explicitly. 

It follows from eq.(\ref{S2}) that $c_{\nu}$ obeys 
\begin{equation}
\partial^2 c_{\nu} = 
{A_0\over 2} \left( \partial^2 h_{\nu 0} + \partial^{\mu} \partial_{\nu}
h_{\mu 0} - \partial^{\mu} \partial_{0} h_{\mu\nu} \right). 
\label{eq-c}
\end{equation}
Therefore, $c\sim A_0h$, so the Lagrangian in eq.(\ref{S2}) is of
order $A_0^2 (\partial h)^2$.

Let us now consider gravitational field of an external source
(ordinary matter). Upon excluding $c_{\mu}$ from the field equations
through eq.(\ref{eq-c}), one obtains an equation for $h_{\mu\nu}$ which
has the following structure, 
\begin{equation}
M^2  \partial^2 h + \epsilon_{vac} h +  A_0^2 \partial^2 h =
T^{ext}. 
\label{schematic-eq}
\end{equation}
where $T_{\mu\nu}^{ext}$ is the external energy-momentum tensor. As we
consider relatively large momenta, $\partial^2 h \gg H^2 h$,
eq.(\ref{main-eq}) implies that the second term here is negligible as
compared to the third one. 

At $M^2\gg A_0^2$, eq.(\ref{schematic-eq}) is the usual weak field
limit of the Einstein equations. However, the Newton gravity constant
is wrong, because $M\gg M_{Pl}$ (see eq.(\ref{AggM})). Hence, we are
lead to consider the opposite case, $A_0\gg M$. Neglecting the first
two terms in eq.(\ref{schematic-eq}) we find that it has the following
explicit form,
\begin{equation}
T^A_{\mu\nu} + T^{ext}_{\mu\nu} = 0
\label{T+Text=0}
\end{equation}
where 
\[
T^A_{\mu\nu} = {\eta\over 2} \biggr\{ 
A_0^2 \Bigl[ g_{\mu 0} ( - \partial^2 h_{\nu 0} - 
\partial_{\nu} \partial^{\lambda}h_{\lambda 0} + 
\partial_{0} \partial^{\lambda}h_{\lambda \nu} ) 
\]
\[
+ g_{\nu 0} ( - \partial^2 h_{\mu 0} - 
\partial_{\mu} \partial^{\lambda}h_{\lambda 0} + 
\partial_{0} \partial^{\lambda}h_{\lambda \mu} ) 
+ \partial_{\mu} \partial_0 h_{\nu 0} + 
\partial_{\nu} \partial_0 h_{\mu 0} 
- \partial_0^2 h_{\mu\nu} \Bigr]
\]
\begin{equation}
+ A_0\Bigl[ g_{\mu 0} ( \partial^2 c_{\nu} + \partial^{\lambda}
\partial_{\nu} c_{\lambda} ) 
+  g_{\nu 0} ( \partial^2 c_{\mu} + \partial^{\lambda}
\partial_{\mu} c_{\lambda} ) 
- \partial_{\mu} \partial_0 c_{\nu} 
- \partial_{\nu} \partial_0 c_{\mu} \Bigr] \biggr\}
\label{T1}
\end{equation}
In the case of static source, the perturbations $c_{\nu}$ and
$h_{\mu\nu}$ can be chosen time-independent, and we further impose
the Coulomb gauge condition, $\partial_i h_{i0} =0$. Then one finds
from eq.(\ref{eq-c}) that $c_0 = (A_0/2) h_{00}$, and $00$-component
of eq.(\ref{T1}) becomes
\[
{\eta\over 2} A_0^2 \partial_i^2 h_{00} + T^{ext}_{00} =0. 
\]
This is the Newton's gravity law (or anti-gravity law, depending on
the sign of $\eta$), but again with the wrong gravitational
constant. We conclude that the effective Planck mass is presently too
large in the Universe evolving according to eqs.(\ref{A0}) and
(\ref{H}).

\section{Modified gravitational waves} 

While the above argument by itself demonstrates that the mechanism of
refs.\cite{Dolgov:1991zc,Dolgov:1997zg} is not viable, there is yet
another undesirable feature of this mechanism which illustrates the
danger of strong breaking of Lorentz invariance. Indeed, let us
consider gravitational waves in this model. These are conveniently
analyzed in the gauge $c_0=0$, $h_{0i}=0$ (the fact that these gauge
conditions may indeed be imposed can be seen by inspection of the
transformation laws of $c_{\mu}$ and $h_{\mu\nu}$ under general
coordinate transformations).  In this gauge eq.(\ref{eq-c}) becomes
\begin{eqnarray}
\nonumber
\partial^2 h_{00} &=& 0, \\
\nonumber 
\partial^2 c_i &=& {A_0\over 2} ( \partial_i \partial_0 h_{00} 
+ \partial_j \partial_0 h_{ij} ) ,
\end{eqnarray}
while the sourceless equation (\ref{T+Text=0}) reads
\begin{eqnarray}
\nonumber
A_0 \partial_0^2 h_{00} - 2\partial_0\partial_i c_i &=& 0, \\
\nonumber
A_0 \partial_0\partial_j h_{ij} + 2\partial_i\partial_j c_j 
+ \partial_i^2 c_i &=& 0, \\
\nonumber
\partial_0 (A_0\partial_0h_{ij} + \partial_i c_j + \partial_i c_j )
&=& 0.
\end{eqnarray}
This set of equations has only longitudinal and vectorial solutions
for metric perturbations, the latter having a peculiar dispersion law:

Longitudinal ($p_0^2={\mbox{\boldmath $p$}}^2$):
\begin{eqnarray}
\nonumber
h_{ij} &=& p_i p_j d({\mbox{\boldmath $p$}}),\\
\nonumber
h_{00} &=& -p_i^2 d({\mbox{\boldmath $p$}}),\\
\nonumber
c_i &=& - {A_0 \over 2} p_0 p_i d({\mbox{\boldmath $p$}}),
\end{eqnarray}

Vectorial ($p_0^2 =  {\mbox{\boldmath $p$}}^2/2$):
\begin{eqnarray}
\nonumber
h_{ij} &=& p_i d_j({\mbox{\boldmath $p$}}) 
+  p_j d_i({\mbox{\boldmath $p$}}),\\
\nonumber
h_{00} &=& 0,\\
\nonumber
c_i &=& - A_0 p_0 d_i({\mbox{\boldmath $p$}}),
\end{eqnarray}
where $d({\mbox{\boldmath $p$}})$ and $d_i({\mbox{\boldmath $p$}})$
are arbitrary amplitudes satisfying the only condition $p_i
d_i=0$. These solutions are drastically different from transverse
traceless gravitational waves of general relativity, and their
emission almost certainly will not pass observational tests.

\section{Discussion}

The problem with Newton's gravity law which rules out the mechanism of
refs.\cite{Dolgov:1991zc,Dolgov:1997zg} is analogous to the
modification of gravity inherent in compensation mechanisms invoking
scalar fields \cite{scalar1}. There, too, the effective Planck mass is
unacceptably large at the present epoch. In addition to this problem,
models with vector and/or tensor compensating fields lead to
unconventional properties of gravitational waves; in particular, these
do not necessarily travel with the speed of light because of the
broken Lorentz invariance.

It appears that undesirable modification of gravity is a generic
property of the mechanisms aimed at compensating the cosmological
constant: all models proposed so far are ruled out because of this
property. It remains to be understood whether there is a way out of
this difficulty, and whether realistic compensation models may be
constructed.

\section*{Acknowledgments}

The authors are indebted to A.~Dolgov for helpful
correspondence. This work is supported in part by Russian Foundation
for Basic Research grant 99-02-18410.

\end{document}